# Semantic based model of Conceptual Work Products for formal verification of complex interactive systems

Mohcine Madkour[1*], Keith Butler[2], Eric Mercer[3], Ali Bahrami[4], Cui Tao[1]

[1] The University of Texas Health Science Center at Houston, School of Biomedical Informatics, 7000 Fannin St Suite 600, Houston, TX 77030

[2] University of Washington, Department of Human Centered Design and Engineering Box 352315, Sieg Hall, Room 208 Seattle, WA 98195

[3] Brigham Young University Computer Science Department, 3334 TMCB PO Box 26576 Provo, UT 84602-6576

[4] Medico System Inc. 10900 NE 8th Street Suite 900 Bellevue, WA 98004

[*] Corresponding author, email: mohcine.madkour@gmail.com, phone: (+1) 281-652-7118

***Abstract*** **- Many clinical workflows depend on interactive computer systems for highly technical, conceptual work products, such as diagnoses, treatment plans, care coordination, and case management. We describe an automatic logic reasoner to verify objective specifications for these highly technical, but abstract, work products that are essential to care. The conceptual work products specifications serve as a fundamental output requirement, which must be clearly stated, correct and solvable. There is strategic importance for such specifications because, in turn, they enable system model checking to verify that machine functions taken with user procedures are actually able to achieve these abstract products. We chose case management of Multiple Sclerosis (MS) outpatients as our use case for its challenging complexity. As a first step, we illustrate how graphical class and state diagrams from UML can be developed and critiqued with subject matter experts to serve as specifications of the conceptual work product of case management. A key feature is that the specification must be declarative and thus independent of any process or technology. UML can represent the needed static and dynamic abstractions but it also allows inconsistent, unsolvable models. Our Work Domain Ontology with tools from Semantic Web is needed to translate UML class and state diagrams for verification of solvability with automatic reasoning. The solvable model will then be ready for subsequent use with model checking on the system of human procedures and machine functions. We used the expressive rule language SPARQL Inferencing Notation (SPIN) to develop formal representations of the UML class diagram, the state machine, and their interactions. Using SPIN, we proved the consistency of the interactions of static and dynamic concepts. We discuss how the new SPIN rule engine could be incorporated in the Object Management Group (OMG)'s Ontology Definition Metamodel (ODM).**

## 1. Introduction

Many critical systems require highly complex user interactions and a large number of cognitive tasks. These systems are common in clinical health care and also many other industries where the consequences of failure can be very expensive or risky to human safety. Formal verification through model checking could reduce or prevent system failures, but several technical obstacles must be solved first.

We focus here on the abstract products of conceptual work that are foundational requirements that must be part of verification in a modern health care systems. Examples of such conceptual work products include diagnoses, medication reconciliation, treatment plans of orders, care coordination, case management, etc. There are several reasons why these conceptual work products are difficult for system developers. They are complex and abstract, with no required tangible manifestation until they are acted upon. They are often developed in an environment of distributed cognition, where no single agent possesses complete knowledge of their state. They are commonly carried out by multiple clinicians with information that must be integrated from computer systems and the physical environment.

Despite their importance and technical content, conceptual work products in health care are often only vaguely defined. Software developers often try define them in terms of user procedures and software features that will be used to create them. This causes a logical problem: in order to serve as a verification criterion, the specification must be stated independently from the system needing verification, or it risks being a tautology instead of an evaluation. More generally, models of information systems that lack clear, yet consistent specifications for their work products are seriously incomplete. This gap forces developers to rely on intuition about this fundamental purpose of a system, even when the domains are highly technical and critical system failures can risk the health, safety or security of large groups of people.

### Declarative Specifications

Declarative models of conceptual work products specify what a system must output in a manner that is independent of how the system will do it. Despite their novelty for interactive systems there are many well established examples for declarative specifications in other industries, such as manufacturing, where declarative specifications are common for physical parts that a manufacturing system must produce. For a simple example, the mounting brackets for airliner seats can be specified in terms of their geometrical shape, their strength, and their weight. Thus, the product specification is a foundational requirement for any system design- it must be capable to produce it. Such declarative specifications state clearly what a manufacturing system must produce in a manner that is independent from how it will be done. Technically, the brackets could be manufactured in several optional ways: by molding from liquid metal, sculpting them from a solid piece, cutting them out with a stamping machine, or even building them with a 3-D printer. The independence of the bracket specification allows systems engineers considerable latitude to analyze the costs and benefits of different technologies to produce equivalent products. If, however, a system cannot meet this fundamental requirement then it will fail to achieve its primary purpose, regardless of how impressive its other features may be. Speaking generally, any system model that does not have clear specification of input and output is seriously incomplete.

## Conceptual Work Products

Models of conceptual work products play an analogous role to manufacturing product specifications. They provide an objective way to define abstract entities of knowledge work that a system must accomplish. They complement procedural models and physical work entities to make a workflow model of modern health care far more complete.

Conceptual work products exploit recent developments in declarative modeling techniques to define them in terms of classes, attributes, relationships, states and transition rules. The transition rules are defined by combinations of attribute values. The workflow of tasks must be capable of changing the values of attributes from the starting state of the arriving entity, through any required intermediate states, to the goal state that is assigned to the system. The class models and state machine models serve as objective, verifiable specifications of the abstract products of knowledge work regardless of how it is distributed across the human and computer agents that carry out a workflow.

In addition to health IT [1][2], the principles of conceptual work products have been demonstrated in domains ranging from aircraft scheduling [3], to mission planning for the International Space Station [4] and for online search [5] and many other applications [6]. In more recent work we developed a Work Domain Ontology for Emergency Department (ED-WDO) that faithfully represents the ED work domain independently from any clinical setting, specific technology or environmental variables [7]

## Modeling Conceptual Work Products

Conceptual work products correspond to a "trigger" in Use Cases [8] or to the arriving "instance" that starts a process in the Business Process Modeling Notation [9]. They are a part of the domain ontology for complex systems. Modeling a domain requires participation by subject matter experts, whose main professional responsibilities do not include software technology. They can, however, recognize how it is represented in the graphical notation of UML class and state diagrams well enough to review and critique them.

The "understandability" of UML notation is an important advantage for subject matter experts, but in order to serve as requirements for health IT systems the models must also be "solvable" and internally consistent. Consistency problems are very common in large, complex UML models. Khan [10] found that 49% of a set of 303 published UML metamodels [11] were not well-formed and they are hard to validate. Wilke et al [12] found that 48.5% of the OCL constraints used for expressing the well-formedness of UML in OMG documents are erroneous. The validation problems of structural diagrams are considerable. A number of problems that have been discussed by other researchers include the consistency of UML class diagrams with hierarchy constraints, the reasoning over UML class diagrams, the full satisfiability of UML class diagrams, and the inconsistency management in model driven engineering.

Our goal is to address this methodological gap and translate UML class diagrams and state machines to a semantic web based model. This process can be decomposed into three steps. First, we validate individual concepts: classes, objects, associations, links, domain and range, multiplicity, composition, unique and non-unique associations, ordering, class generalization, and association generalization. Second, we validate states and state transitions invariants. Third, we validate interclass constraints and the cohesion between state transitions constraints and class model constraints. As for the representation of the conceptual work product specification for model checking, we build a semantic based model of state machine that formally describe states and transitions. This model will serve to distinguish different behaviors that can be represented by the model checker. In addition to providing a system-independent specification of conceptual work product, domain ontology has the automatic logic reasoner tool for verifying the logic consistency in objective specifications. This has been considered as a new feature added to declarative knowledge modeling tools to address the gap for UML consistency verification.

Figure 1 below depicts the overall method for verifying interactive systems. It shows the key position of the current research to translate the conceptual work product represented by UML blueprints into an ontology based model in order to verify its logical solvability. The model checker is then called upon to verify the model against the automata.

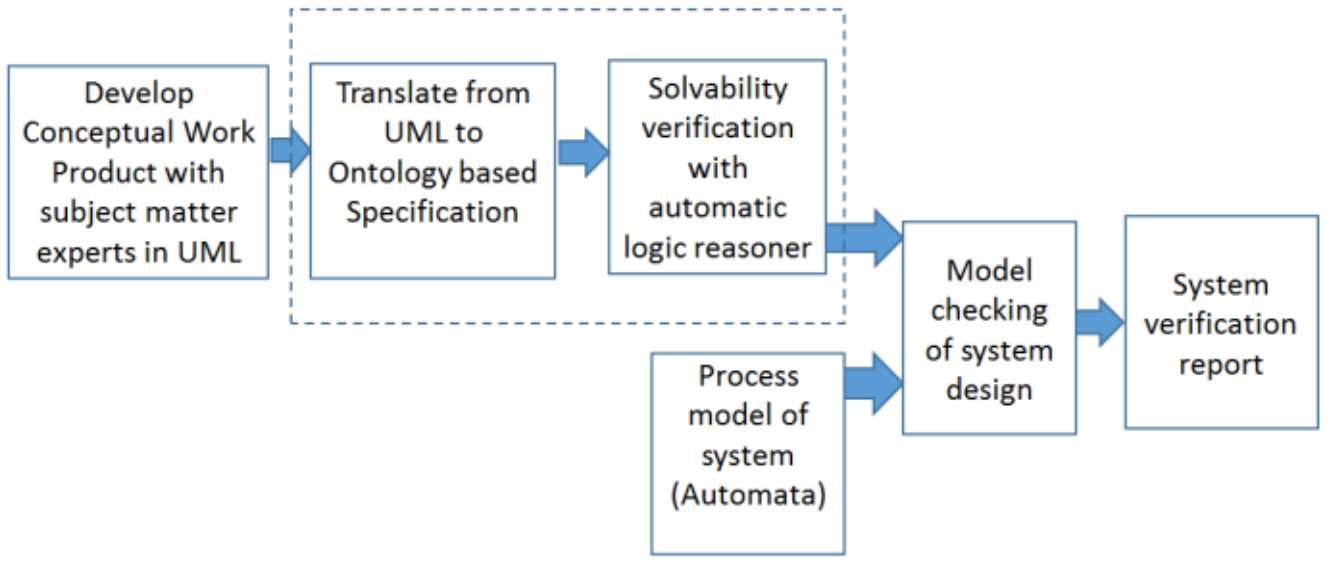

FIGURE 1: FLOW DIAGRAM OF SOLVABILITY VERIFICATION OF CONCEPTUAL WORK PRODUCT USING UML BASED MODEL

The ontology-based representation enables the validation of automatic consistency checking of input data and serves as a formal specification to a model checker. In this work, we describe and provide solutions for the challenges of translating specifications from UML to ontology and we describe the solvability verification method (the two components framed by the dashed rectangle in the Figure 1). We use SPARQL Inferencing Notation (SPIN) [13], a Semantic Web modeling language available as a W3C standard. We demonstrated our propositions with the help of a case management use case, implementing a number of tasks, including assessment, planning, coordination, evaluation, communication and collaboration. The remainder of this paper is structured as follows. Section 2 presents the use case of patient-centered case management. Section 3 presents some concepts mapping between UML and Semantic Web. Section 4 describes the SPARQL Inference Notation based approach to model UML class diagram and state machine. Section 5 discusses and concludes this work.

## 2. An Example of Patient Centered Case Management

To illustrate the specification description of highly complex and critical conceptual work products, we present an example of a patient-centered case management for a Multiple Sclerosis (MS) outpatient care clinic [14]. We use [1] a model-based design method for representing an interactive health information technology system that extends a workflow model with conceptual work products. The work domain ontology helped us, not only to abstract away the complexity introduced by particular information systems and work procedure, but also to provide explicit specifications for the information product they must produce. The case management consists of a number of tasks, including assessment, planning, coordination, and evaluation that involve intensive communication and collaboration plans. It provides the clinic's patients with a single point of contact to maintain situation awareness of each patient's plan and intervene if orders are not carried out correctly. The case management use case example takes place between clinic visits for outpatients with chronic disease and complex conditions. Following each patient encounter, a doctor develops a treatment plan that typically contains several different orders, such as medication prescriptions, blood tests, images and scans, and consults with specialists and therapists. The orders are often carried out over different steps and time courses (e.g., an x-ray can take place the same day as the visit without an appointment, the neurological exam needs an appointment and will happen later in time) and require integration of diverse sources of information.

In this vein, the conceptual work product can provide answers to many questions that may unfold in the case management process such as: What are the orders that need immediate attention. Which patients have treatment plans that are not progressing? Which patients have plans that are beginning to fall behind their progress profile? What can we do next to accelerate the process of care? The conceptual work product specification prescribes what to do and what not. The specifications have required properties that need to be valid at any time; such a property can be that a care plan should never be able to reach a situation in which no progress can be made (a deadlock scenario). The conceptual work product is considered to be "correct" whenever its specifications satisfy all their properties, therefore, correctness is always relative to a specification, and is not an absolute property of a system. Figures 2 and 3 show the UML class diagram and state machine of the MS-clinic Patient-Centered Case management [1]. The treatment plan class is a composition of order, patient-initiated contact, and self-assigned task classes. The case manager checks patient's orders, initiates contacts, and places self-assigned tasks for particular orders. An order can be specified to different classes namely exam, prescription, lab test, equipment order, imaging, and consult. The state machine (also called state diagram) illustrates the behavior of classes' objects in response to a series of defined events that act as internal/external stimuli. Figure 3 shows the state machine of order' objects.

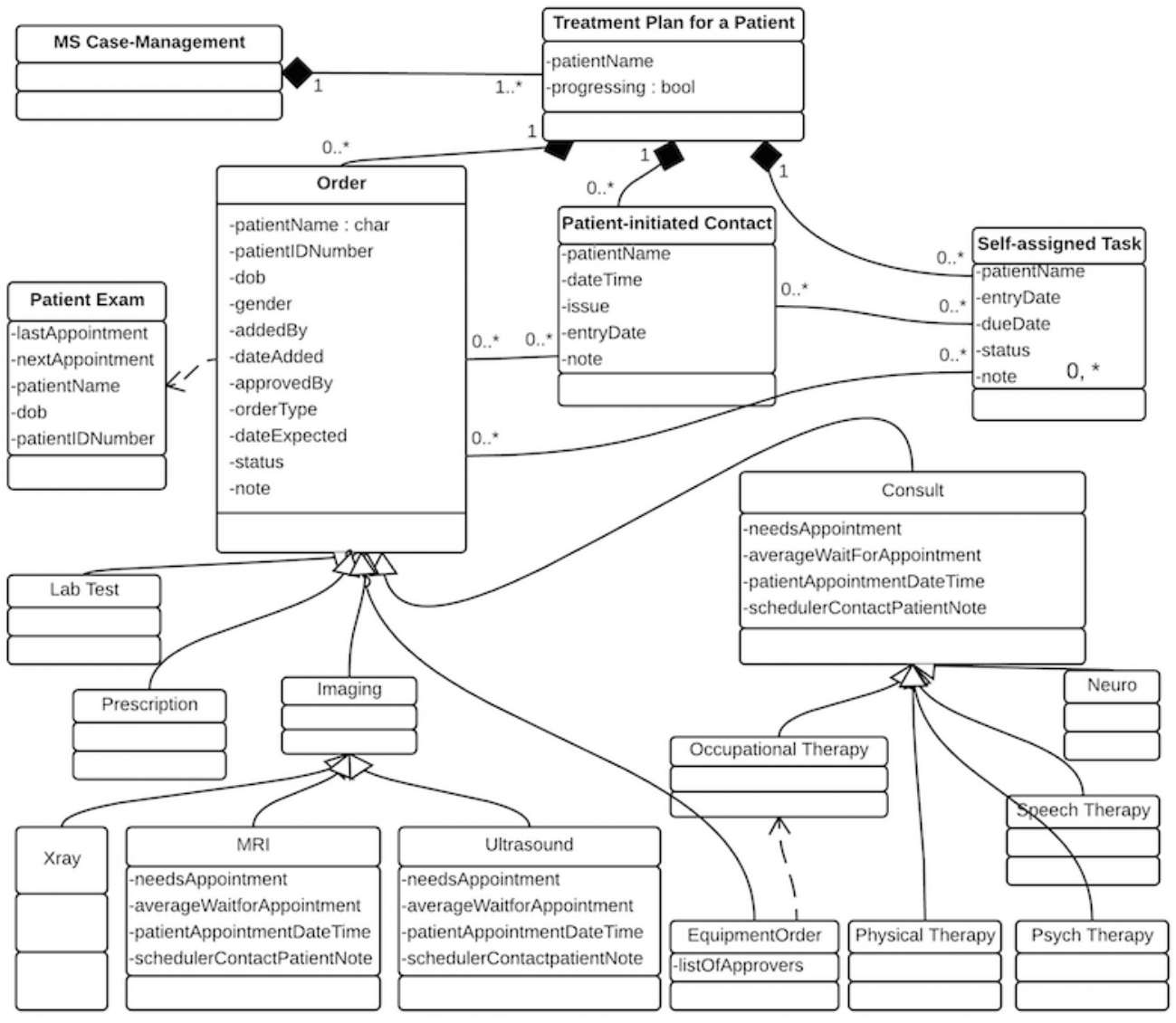

FIGURE 2: UML CLASS DIAGRAM OF PATIENT CENTERED CASE MANAGEMENT

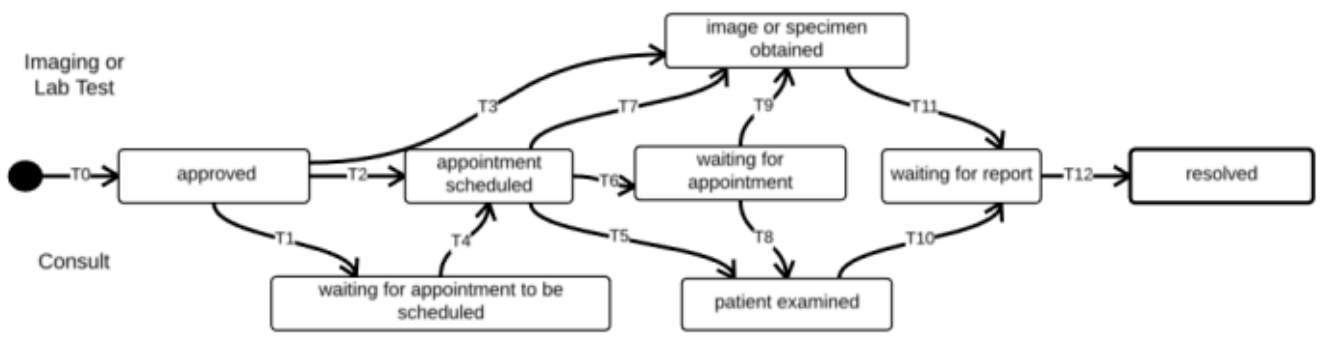

FIGURE 3: UML BASED STATE MACHINE OF OBJECTS OF CLASS

## 3. Mapping UML Model to Semantic Web Based Ontology

In our analysis between OWL and UML modeling, we found that both language definitions refer to comparable meta-models laid down in terms of OMG's Meta Object Facility, but in contrast to UML, OWL is fully built upon formal logic, which allows logical reasoning on OWL ontologies. UML covers considerable ground on the behavioral side, and there are also an increasing number of metamodels and other profiles (e.g., SysML, SoaML, BPMN, etc.) that are relevant to any transformation to OWL. Table 1 summarizes the comparison between the two frameworks. We found that artifacts in UML design do not have similar metamodels of Semantic Web concepts. The aggregation, composition, class methods, associations, and instances, are object-oriented notations and don't have similar functions in the Semantic Web, which is based on rules and inferencing reasoning. In addition, we found that basic concepts used by the two languages have quite similar meanings. Therefore, a class and the attributes of a class in UML are similar to the concept of a class in OWL and OWL Object properties or Datatype properties. OWL properties can be declared with directions (i.e. domain and range), similarly to class attributes in UML and UML associations. However, there are many differences between the two modeling languages. For instance, OWL has the property owl: inverseof to denote inverse property, UML does not have a similar feature. The multiplicity is specified in OWL by the cardinality constraint owl: cardinality, which denotes the number of values a property can have, whereas UML represents multiplicity using lower and upper bounds, which remains inconsistent (i.e. asterisk means unlimited but cannot be infinite).

TABLE1: GENERAL AND SPECIFIC COMPARISON BETWEEN UML AND OWL

| Ontology based model | UML based model |
|---|---|
| Based on the structural properties of a class | Centered around methods on class |
| Oriented towards Object Oriented Analysis (OOA) and classification | Oriented towards Object Oriented Programming (OOP) and Object Oriented Design (OOD) |
| Reasoning on data at runtime | Runtime knowledge is left to database or object in programming language such as C++ and Java |
| Formal language (explicit and precise and subject to mathematical and/or logical reasoning) | Less precise but the concepts specified in the UML are described and used so clearly and explicitly in a common standard way |
| Principle of minimality (built upon formal logic which allows logical reasoning) | Aims at maximal expressivity: 13 Different 'model types' for different aspects of a system |
| An object (individual) may be independent from any class or belong to many classes at the same time | An object can only be of one class at a time |

The Object Management Group (OMG) released their Ontology Definition Metamodel (ODM) specification [15] in order to bridge the gap between Model Driven Architecture (MDA) and the Semantic Web. The ODM represents formal logic languages, such as description logic (DL), common logic (CL) and first-order predicate logic. ODM provides mappings to OWL-DL and a UML2 profile for ontologies. ODM, however, is a standard addressing ontology description, but not reasoning. The reasoning component, which is important in our framework, would need to be addressed in addition to the standard.

### 3.1 Mapping Class, Attriutes, and Associaitons from UML to the Semantic Web Notations

Much previous works discussed the semantic preserving transformations of UML and identified similarities and incompatibilities between UML and ontology languages[16] [17] [18] [19] [20]. It has been found that the concept of Property in DAML+OIL, although similar to the association in UML, cannot be mapped easily, therefore it has been suggested to add a meta class Property for UML to make it compatible with knowledge representation languages and support the aspect-oriented programming. Mehrolhassani et al [21] pointed out that the conversion from UML to OWL is not as straightforward as it seems and proposed some translation rules such as using owl:hasValue to assign a default value to an attribute, rdfs:comments to convert roles that are placed near end of the associations, and properties owl:IntersectionOf, owl:unionOf and owl:complementOf to convert, respectively, "AND", "OR" and "NOT" relationships in UML. Zedlitz et al's work on converting datatypes between UML and OWL proved that the datatype axiom in DL are capable of representing UML user-defined datatypes [22]. The properties DataOneOf and DataUnionOf in OWL have been used to assign range of individuals and enumerate sets of objects and classes for UML class enumeration. The HasKey property is used to restrain each attribute to have only one value. It is used for the conversion of UML datatypes that are similar to classes but their instances of a data type are identified only by their value. Khan et al [23] proposed an automatic tool that implements UML to OWL translation. His approach provides a strict translation of an instance of UML class by using the OWL notions of equivalence and disjointness and proposed NegativeObjectPropertyAssertion for representing the association existing between classes only and not instances. It also translates UML ordered property into object property with many functional and inverse-functional subproperties indexes, each represents a unique number in the sequencing order of the property. In addition, to express unique association in UML, the approach uses ObjectMinCardinality and ObjectMaxCardinality, and for non-unique association, the property is left unconstrained to allow multiple links between the objects of a domain class and a range class.

### 3.2 Translation of the Value-partition and Part-whole UML Associations

The value partition and part-whole relationships are two frequently used association types in class diagram and many approaches have different interpretations when translating them to OWL Ontology. The value-partition (value-set representation) is required when representing class's features in UML class diagram such as class attributes. OWL uses the same concept of object oriented representation in class's attributes representation but the meaninf is slightly different. We found there are two major interpretations: one is to represent attributes as disjoints individuals and the other is to represent them as disjoint classes. The first method derives from a concrete separation of the relation concept-feature and the second uses the notion of continuous space of the class-feature.

For example, the state of a treatment plan object in our use case takes different values that we consider as attributes of a class: progressing state, hung state, approved state, and complete state. Using the first translation approach, the class treatment plan state will be represented as an enumeration (alone) of four disjoint individuals. In this case, no further sub-partitioning of the values is allowed neither accumulating two or more values for defining new value for an object. In the second translation approach, the expression "the plan is progressing" is just as to say "the plan's state is inside the sub-class progressing of the treatment plan state class" which means that a new class for each feature value needs to be created. The latter method enables the possibility to break down a value of a state into smaller sub-partitions. This will allow to create new values such as "slowly progressing" or "quickly progressing" by subdividing the "progressing" value. In addition, comparing to the first representation which uses instances of classes, the second method requires that an instance of a feature class is created each time a feature individual is initiated.

When it comes to translating UML part-whole relationship, OWL does not provide any built-ins for aggregation or composition relations as it does for the direct class-subclass relationship. However there are some already predefined OWL restrictions of the UML aggregation and composition relationships that can be represented. Such notation in UML therefore can be translated in OWL by specifying that a class in OWL cannot be in an aggregation association with itself and a class cannot be a composite of more than one aggregation/composition, and an object of a class that is part of a composition (not aggregation) cannot exist without the class that it belongs to. In cases where there is a hierarchy of compositions and/or aggregation, there are two options to choose from, we can either have different OWL properties defining each UML association instance in the hierarchy, or have the same OWL property definition for all levels of aggregations/composition.

For instance, in Figure 2, the case management class aggregates the treatment plan class which itself aggregates the order class, the self assigned task class, and the patient initiated contact classe. We used the first approach to represent these associations, so we created different OWL object properties in order to convert these associations to OWL, which are "hasPlan", "hasSelfAssignedTask", "hasPatientInitiatedContact" and "hasOrder". In addition, we attributed these object properties the description logic axioms irreflexive (IrreflexiveObjectProperty) and inverse-functional (InverseFunctionalProperty), and as those aggregations are of type composition we denote their inverse properties: "planOf", "orderOf", "patientInitiatedContactOf", and "selfAssignedTaskOf" with cardinality restrictions set to one. However, using this approach, these OWL object properties remain unrelated, therefore, in order to assure the transitivity of aggregation relationship, for example between (hasPlan – hasOrder) , (hasPlan –hasSelfAssignedTask), and (hasPlan – hasPatientInitiatedContact), we use the description logic axiom OWL Object Property chain.

The second alternative is to define a single relation (e.g. hasPart) for all composition and aggregation associations. The hasPart relation will have an inverse property (e.g. partOf ) with the OWL axiom TransitiveObjectProperty and should not have any range or domain restrictions so that the composition/aggregation property can be used in different positions in the ontology. To enforce the semantic of the whole-part relation, one can define allValuesFrom restriction in the classes to constraint that they must have some parts (partOf) or must be parts of (hasPart) some specified classes. Moreover, the requirement to restrict the relation hasPart to have the inverse functional characteristic is not possible because OWL does not enable transitive properties to have any cardinality restriction.

Up to now the behavior notations in OWL that represent further constraint on OWL such as those expressed in the integrity checking in UML class diagram are missing. This is crucial for accomplishing the translating of large and complex conceptual work products that have many classes and relationships and that need many constraints to represent the defined concepts.

## 4. SPARQL Inference Notation based Approach for Representing UML Class Diagram and State machine

The Semantic Web offers a formal framework for modeling conceptual work products and a comprehensive modeling technology stack that consists of a suite of related standards including Resource Description Framework (RDF), OWL, SPARQL (the query language of the semantic web), and SPARQL Inference Notation (SPIN). OWL uses Description Logic to define semantic axioms, RDF provides a general approach for describing data of any kind, and SPIN is designed to represent SPARQL rules and constraints. However, with the growing complexity of conceptual work products, it turns out that OWL description logic axioms support only simple constraints and cannot model the behavior of objects as described in UML behavioral diagrams such as state machine. We use SPIN framework to provide a semantic based approach to represent conceptual work products because of its rich expressivity, consistency checking, and automatic model validation capabilities. SPIN provides meta-modeling capabilities that allow users to define new functions and query templates and has a ready to use library of common functions (i.e. common functions, complex mathematical calculations, modules and templates) that can be used to instantiate re-usable SPARQL queries in RDF and OWL ontologies to add rules and constraints checking to the ontologies. In addition, SPIN provides an automatically deterministic reasoning over defined concepts and rules. The SPIN constraint checking engine verifies instantaneously and systematically the consistency of the data with the rules and raises warnings when a violation of constraint occurs. Using SPIN, constraints, rules and computable concepts that form a conceptual work product can be stored in RDF data format in an ontology-based model. We used the SPIN editor TopBraid Composer [24] to illustrate the representation of conceptual work products in the case management use case . The editor provides a user interface for SPIN based modeling and ontology engineering [25][26].

### 4.1. Semantic based Representation and Validation of UML class diagram

Figure 4 illustrates the SPIN based class diagram for the UML class diagram of case management use case. The diagram consists of owl: classes, owl:ObjectProperties, and owl:DatatypeProperties that we created for translating the UML clases, UML attributes, and UML associations. The OWL Object property links two resources within the ontology, the OWL data type property links resources to their data types. These properties are used to define domains and ranges restrictions of classes. Other used types of restrictions used include quantifier restrictions such as allValuesFrom and someValuesFrom and cardinality restrictions such as minCardinality, cardinality, maxCardinality, and hasValue restrictions. These are used to restrict the individuals that belong to a class. More complex constraints on conceptual work products can be represented by SPARQL queries. SPARQL provides five different query variations which can be used for different purposes, namely the SELECT query, CONSTRUCT query, ASK query, DESCRIBE query, and UPDATE query. We use the ASK WHERE query type to represent constraints and the CONSTRUCT query type to infer new RDF triples about objects of conceptual work products. SPARQL Inference Notation defines three levels of manipulating data, namely, CONSTRAINTS to verify data, CONSTRUCT to infer data and RULE to search and update data.

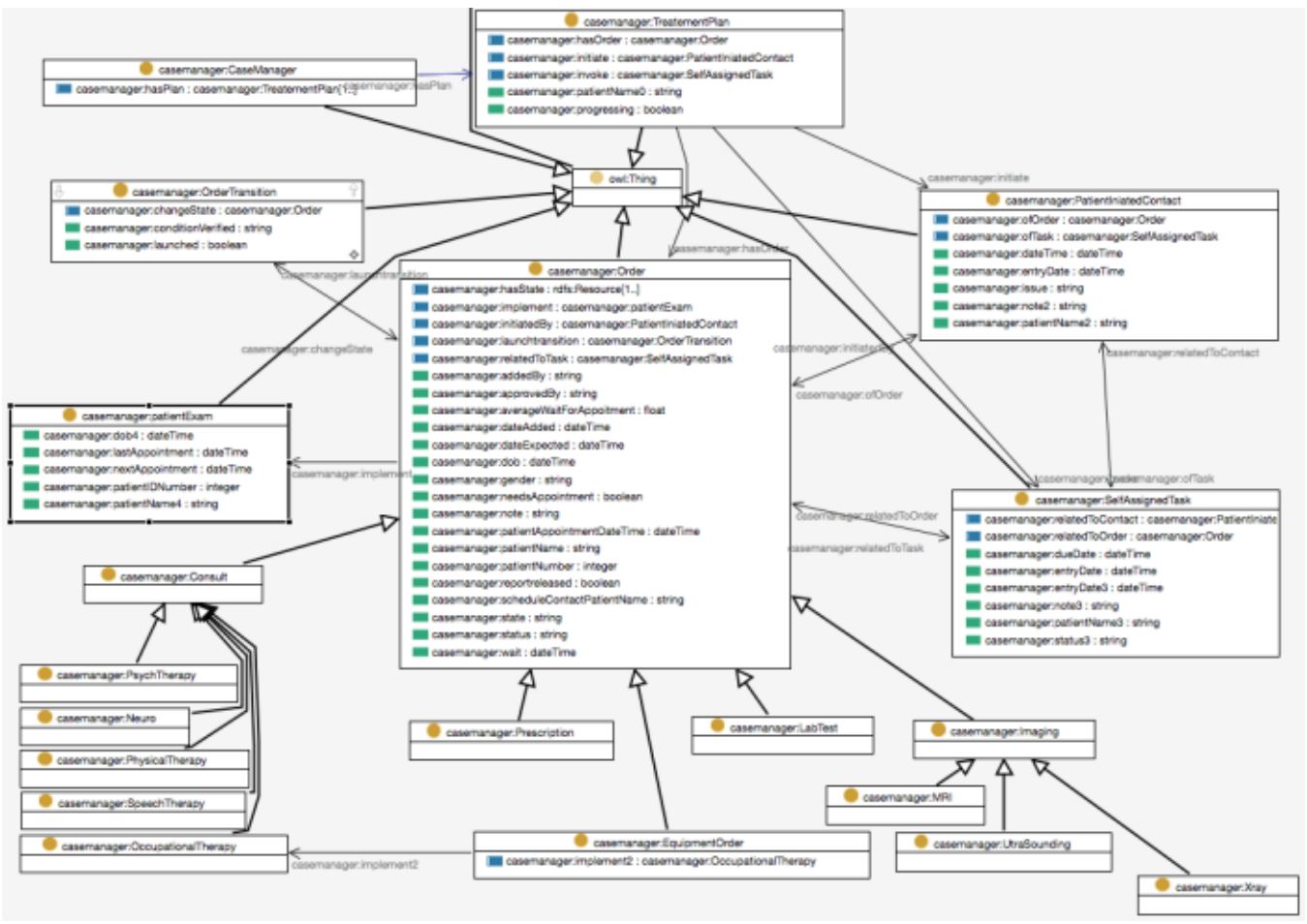

FIGURE 4: SPARQL INFERENCE NOTATION BASED CLASS DIAGRAM OF UML CLASS DIAGRAM

The rules in Figure 5 show seven constraints defining the class Order which are represented in ASK constraint queries. Each of these queries takes a WHERE block to restrict the query. The FILTER eliminates solutions that do not cause an expression to evaluate to true. The first constraint checks the gender of the patient; so, each instance of Order must have either male or female in gender property. The second and third constraints are inter-class relations, they check on the patientName property in class Order to verify if it matches with patientName in SelfAssignedTask and PatientInitiatedContact. The forth constraint describes that instances of Order must have expectedDate greater than addedDate. The fifth constraint specifies that patientNumberID must be greater than 0. The last two constraints in the figure 5 represent the composition relationship between TreatmentPlan and Order by specifying that order can only belong to one plan. The NOT EXIST restricts an order to be attached to a treatment plan using deterministic reasoning of closed semantics of the rule-based system. The last example in Figure 5 represents a CONSTRUCT query based rule to initialize the state of new Order instances to “initial” value. Thus whenever a binding of the pattern in the WHERE clause occurs, the RDF triple in the CONSTRUCT clause is inferred and added to the concurrent RDF dataset.

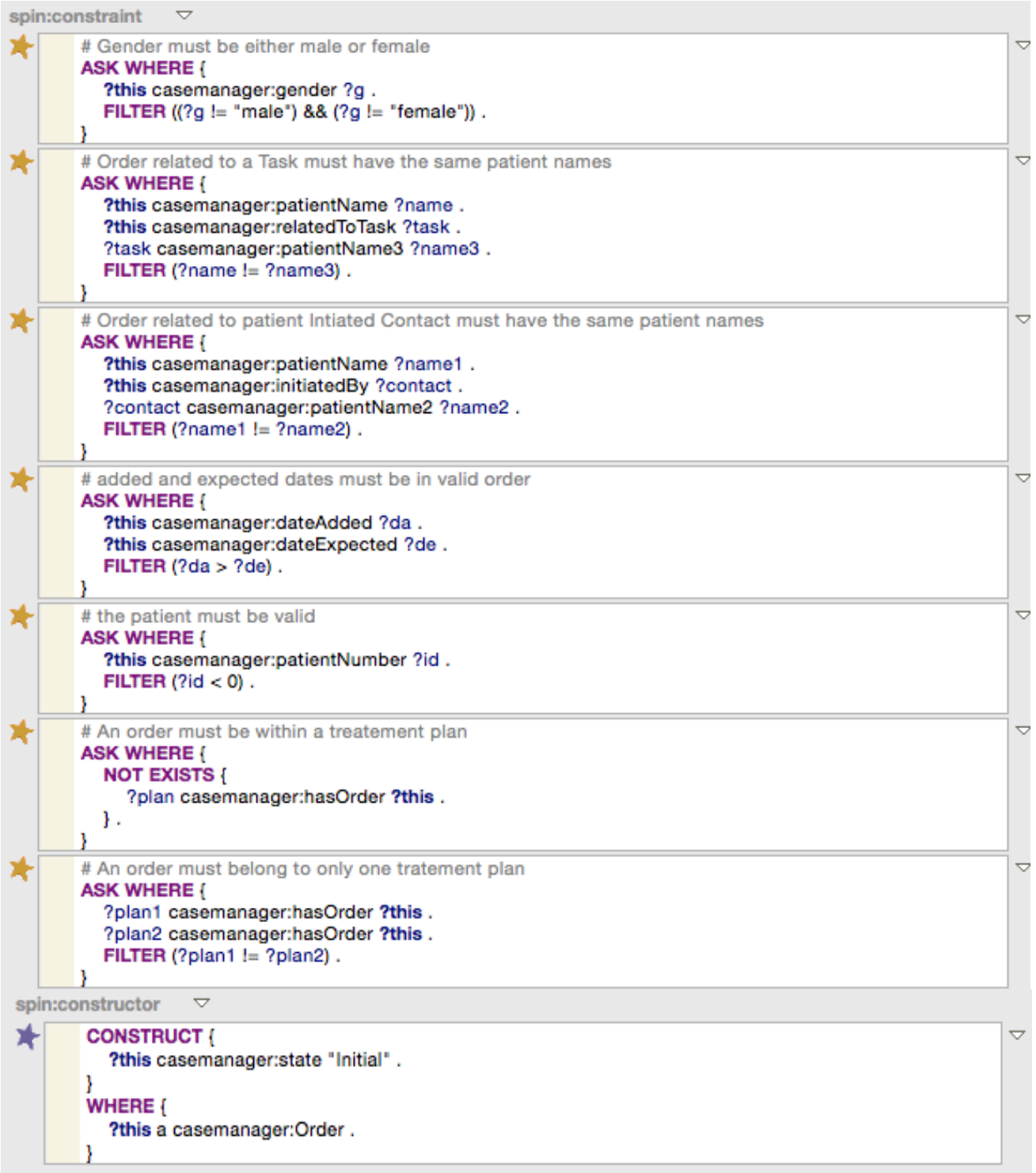

FIGURE 5: ILLUSTRATION OF ASK CONSTRAINS TO DESCRIBE ORDER OBJECTS AND CONSTRUCT TO INITIALIZE ORDERS INITIAL STATES IN SPARQL INFERENCE NOTATION

### 4.2. Semantic Based Representation and Validation of State Machine

The visual formalisms of the UML state machine have been used for enabling modular conceptual modeling of complex systems. Therefore many alternatives for its formal representation have been proposed for formal model checking [27][28] [29]. The momentum gained by state machines is due to their role in formal specification of behaviors in critical systems (i.e. events, conditions, actions, and constraints). A state in state machine is defined by a set of invariants whose values hold unchanged. A transition from one state to another is verified by guard conditions, and consists of a set of constraints and rules. Table 2 shows the conditions of transitions for order objects in the state machine of the case management use case. The transitions are enumerated from 1 to 12 as they appear in the state machine in Figure 3. The properties used in guard conditions are defined in classes definitions (Figure 4).

Table 2: Transitions of order's states

| Transition | Guard conditions |
|---|---|
| T0 | Doctor has approved the order and the valid entry and expected dates have been entered (optionally). |
| T1 | The order needs an appointment and the date of appointment is not decided yet. |
| T2 | The order needs an appointment and a date has been fixed |
| T3 | The imaging or lab test order doesn't need appointment and the image or specimen has been obtained. |
| T4 | A date of the appointment has been fixed for the order consult. |
| T5 | The examination is done for the order consult. |
| T6 | The appointment date is in the future. |
| T7 | The image or specimen has been obtained. |
| T8 | The order consult is done. |
| T9 | The image or specimen is obtained. |
| T10 | The report of the image or specimen is pending. |
| T11 | The report of the consult is pending. |
| T12 | The report of the consult has been released. |

Figure 6 illustrates the representation in SPARQL Inference Notation of the classes Order and OrderTransition connected by the properties casemanager:launchtransition and casemanager:changeState. The state value is represented as owl:DatatypeProperties in class order and we use a transition guard represented by a string datatype called conditionVerified and a boolean data properties called launched to control the transitions between states.

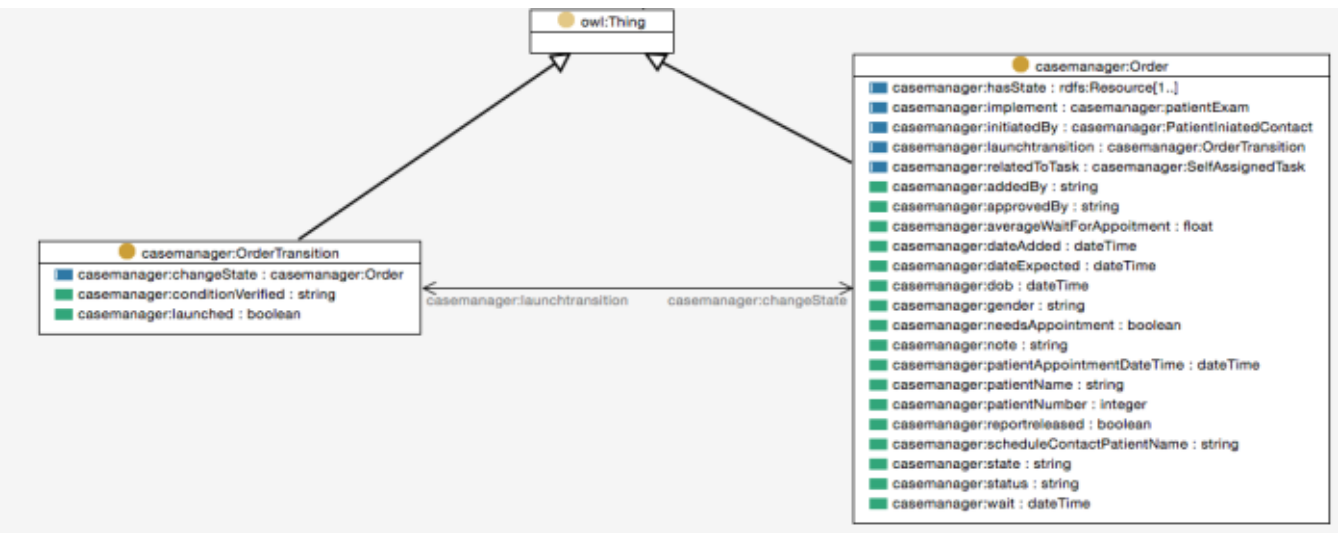

FIGURE 6: TRANSITION AND ORDER CLASSES REPRESENTATIONS IN SPARQL

As represented in the rule in Figure 5, an order is set to the state "initial" at the start and reaches the state "resolved" as the final state. An order can be specified to different class types, namely consult, imaging, and lab test. These classes may have different states and transitions. For example, an order of type lab test or imaging do not have the state "Waiting for appointment to be scheduled" and "Patient examined" states rather they have the state "Image or Specimen Obtained". Figure 7 illustrates the SPARQL rules edited on TopBraid composer to model the transitions in the state machine of objects of type order. The SPARQL RULE has the pattern DELETE {} INSERT {} Where {} which has a delete phase in clause DELETE, and an assertion phase in the clause INSERT and takes place when the condition in the WHERE clause is verified.

```
# T0
DELETE {
  ?o casemanager:state "Initial" .
  ?this casemanager:conditionVerified ?oldCond .
}
INSERT {
  ?o casemanager:state "Approved" .
  ?this casemanager:conditionVerified "valide entry and expected dates" .
}
WHERE {
  ?this casemanager:changeState ?o .
  ?o casemanager:dateAdded ?dateadd .
  ?o casemanager:dateExpected ?dateexp .
  ?o casemanager:approvedBy ?doctor .
  ?o casemanager:state "Initial" .
}
```

```
# T1
DELETE {
  ?o casemanager:state "Approved" .
  ?this casemanager:conditionVerified ?oldCond .
}
INSERT {
  ?o casemanager:state "Waiting for appointment to be scheduled" .
  ?this casemanager:conditionVerified "needs appiotment which not yet assigned" .
}
WHERE {
  ?this casemanager:changeState ?o .
  ?o casemanager:needsAppointment true .
  NOT EXISTS {
    ?o casemanager:patientAppointmentDateTime ?dt .
  } .
  ?o casemanager:state "Approved" .
}
```

```
# T2
DELETE {
  ?o casemanager:state "Approved" .
  ?this casemanager:conditionVerified ?oldCond .
}
INSERT {
  ?o casemanager:state "Appointment scheduled" .
  ?this casemanager:conditionVerified "needs appointment that is assigned" .
}
WHERE {
  ?this casemanager:changeState ?o .
  ?o casemanager:needsAppointment true .
  EXISTS {
    ?o casemanager:patientAppointmentDateTime ?dt .
  } .
  ?o casemanager:state "Approved" .
}
```

```
# T3
DELETE {
  ?o casemanager:state "Approved" .
  ?this casemanager:conditionVerified ?oldCond .
}
INSERT {
  ?o casemanager:state "Image or specimen obtained" .
  ?this casemanager:conditionVerified "do not need any appointment and its already done" .
}
WHERE {
  ?this casemanager:changeState ?o .
  ?o casemanager:needsAppointment false .
  ?o casemanager:status "done" .
  ?o a casemanager:LabTest .
  ?o casemanager:state "Approved" .
}
```

```
# T3'
DELETE {
  ?o casemanager:state "Approved" .
  ?this casemanager:conditionVerified ?oldCond .
}
INSERT {
  ?o casemanager:state "Image or specimen obtained" .
  ?this casemanager:conditionVerified "do not need any appointment and its already done" .
}
WHERE {
  ?this casemanager:changeState ?o .
  ?o casemanager:needsAppointment false .
  ?o casemanager:status "done" .
  ?o a casemanager:Imaging .
  ?o casemanager:state "Approved" .
}
```

```
# T4
DELETE {
  ?o casemanager:state "Waiting for appointment to be scheduled" .
  ?this casemanager:conditionVerified ?oldCond .
}
INSERT {
  ?o casemanager:state "Appointment scheduled" .
  ?this casemanager:conditionVerified "patientAppointmentDateTime assigned"
}
WHERE {
  ?this casemanager:changeState ?o .
  ?o casemanager:patientAppointmentDateTime ?appdate .
  ?o casemanager:state "Waiting for appointment to be scheduled" .
}
```

```
# T5
DELETE {
  ?o casemanager:state "Appointment scheduled" .
  ?this casemanager:conditionVerified ?oldCond .
}
INSERT {
  ?o casemanager:state "Patient examined" .
  ?this casemanager:conditionVerified "examination done" .
}
WHERE {
  ?this casemanager:changeState ?o .
  ?o a casemanager:Consult .
  ?o casemanager:status "done" .
  ?o casemanager:state "Appointment scheduled" .
}
```

```
# T6
DELETE {
  ?o casemanager:state "Appointment scheduled" .
  ?this casemanager:conditionVerified ?oldCond .
}
INSERT {
  ?o casemanager:state "Waiting for appointment" .
  ?this casemanager:conditionVerified "appointment time has not yet arrived" .
}
WHERE {
  ?this casemanager:changeState ?o .
  ?o casemanager:patientAppointmentDateTime ?appdate .
  ?o casemanager:state "Appointment scheduled" .
  NOT EXISTS {
    ?o casemanager:status "done" .
  } .
  FILTER (?appdate < now()) .
}
```

```
# T7
DELETE {
  ?o casemanager:state "Appointment scheduled" .
  ?this casemanager:conditionVerified ?oldCond .
}
INSERT {
  ?o casemanager:state "Image or specimen obtained" .
  ?this casemanager:conditionVerified "examination done" .
}
WHERE {
  ?this casemanager:changeState ?o .
  ?o a casemanager:Imaging .
  ?o casemanager:status "done" .
  ?o casemanager:state "Appointment scheduled" .
}
```

```
# T7'
DELETE {
  ?o casemanager:state "Appointment scheduled" .
  ?this casemanager:conditionVerified ?oldCond .
}
INSERT {
  ?o casemanager:state "Image or specimen obtained" .
  ?this casemanager:conditionVerified "examination done" .
}
WHERE {
  ?this casemanager:changeState ?o .
  ?o a casemanager:LabTest .
  ?o casemanager:status "done" .
  ?o casemanager:state "Appointment scheduled" .
}
```

```
# T8
DELETE {
  ?o casemanager:state "Waiting for appointment" .
  ?this casemanager:conditionVerified ?oldCond .
}
INSERT {
  ?o casemanager:state "Patient examined" .
  ?this casemanager:conditionVerified "examination done" .
}
WHERE {
  ?this casemanager:changeState ?o .
  ?o a casemanager:Consult .
  ?o casemanager:status "done" .
  ?o casemanager:state "Waiting for appointment" .
}
```

```
# T9
DELETE {
  ?o casemanager:state "Waiting for appointment" .
  ?this casemanager:conditionVerified ?oldCond .
}
INSERT {
  ?o casemanager:state "Image or specimen obtained" .
  ?this casemanager:conditionVerified "examination done" .
}
WHERE {
  ?this casemanager:changeState ?o .
  ?o a casemanager:Imaging .
  ?o casemanager:status "done" .
  ?o casemanager:state "Waiting for appointement" .
}
```

```
# T9'
DELETE {
   ?o casemanager:state "Waiting for appointment" .
   ?this casemanager:conditionVerified ?oldCond .
}
INSERT {
   ?o casemanager:state "Image or specimen obtained" .
   ?this casemanager:conditionVerified "examination done" .
}
WHERE {
   ?this casemanager:changeState ?o .
   ?o a casemanager:LabTest .
   ?o casemanager:status "done" .
   ?o casemanager:state "Waiting for appointement" .
}
```

```
# T10
DELETE {
   ?o casemanager:state "Patient examined" .
   ?this casemanager:conditionVerified ?oldCond .
}
INSERT {
   ?o casemanager:state "Waiting for report" .
   ?this casemanager:conditionVerified "report not yet released" .
}
WHERE {
   ?this casemanager:changeState ?o .
   ?o a casemanager:Consult .
   ?o casemanager:reportreleased false .
   ?o casemanager:state "Patient examined" .
}
```

```
# T11
DELETE {
   ?o casemanager:state "Image or specimen obtained" .
   ?this casemanager:conditionVerified ?oldCond .
}
INSERT {
   ?this casemanager:conditionVerified "report not yet released" .
   ?o casemanager:state "Waiting for report" .
}
WHERE {
   ?this casemanager:changeState ?o .
   ?o a casemanager:Imaging .
   ?o casemanager:reportreleased false .
   ?o casemanager:state "Image or specimen obtained" .
}
```

```
# T11'
DELETE {
   ?o casemanager:state "Image or specimen obtained" .
   ?this casemanager:conditionVerified ?oldCond .
}
INSERT {
   ?this casemanager:conditionVerified "report not yet released" .
   ?o casemanager:state "Waiting for report" .
}
WHERE {
   ?this casemanager:changeState ?o .
   ?o a casemanager:LabTest .
   ?o casemanager:reportreleased false .
   ?o casemanager:state "Image or specimen obtained" .
}
```

```
# T12
DELETE {
   ?o casemanager:state "Waiting for report" .
   ?this casemanager:conditionVerified ?oldCond .
}
INSERT {
   ?o casemanager:state "Resolved" .
   ?this casemanager:conditionVerified "report has been released" .
}
WHERE {
   ?this casemanager:changeState ?o .
   ?o casemanager:reportreleased true .
   ?o casemanager:state "Waiting for report" .
}
```

FIGURE 7 SPARQL RULES FOR OBJECT BASED STATE MACHINE TRANSITIONS

The SPARQL rules have been evaluated through a scenario simulation of objects of class order. The ultimate challenge was to fire only adequate rules when changes happen, the results were satisfactory and the rule set is validated

## 5. Discussion and Conclusion

This paper demonstrates the proof of concept of an ontology-based technique to provide formal specifications of conceptual work products of interactive systems. Our approach enables automatic consistency checking based on UML-style visual modelling of conceptual work products and through mapping to SPIN modeling language. We instantiated reusable SPARQL queries in RDF and OWL ontologies to add inference rules and constraint checking to the ontology model. We are among the first to use SPIN for translation of UML conceptual models. Our experience in doing this work confirms the extensibility and flexibility of this approach to large, complex conceptual work products.

Our approach has strategic value to verify interactive systems, and more generally to OMG's Ontology Definition Metamodel (ODM). The OMG has recognized the importance of logic-supported semantic modelling using ontologies, which is reflected in the OMG's ODM initiative. The OMG supports selected methods and platforms through an adoption process. While the formal specification of conceptual work products of interactive systems has not been adopted so far, the need for a specific ODM solution for the verification of complex specifications of conceptual work products is of concern. The current OMG's ODM initiative to define and standardize ontology metamodel will allow the integration of our framework with OMG standards

Our solution builds on existing tools of declarative knowledge modeling for representing specifications of complex conceptual work products. The UML behavioral state machine specifies discrete behavior of a part of a designed system through finite state transitions. The two essential features behind this task are the in-depth description of states and orthogonality of transitions. The in-depth description is the ability of moving back and forth between levels of abstractions of states by clustering and refinement in states (i.e. zooming-in and –out). These two properties are fulfilled by the OR and XOR relationships and they allow consideration of different levels/cuts in the behavior of an object. The orthogonality of transitions is the ability to decompose the transitions into sets of components that can operate synchronously and independently. This property is fulfilled by AND decomposition in state machine. It is proved useful in avoiding blowups on the states because the semantic of orthogonally permits to concurrently and independently share states and properties between independent components.

The SPIN used in this approach augmented the conceptual modeling of UML class diagram and state machine by providing the required consistency checking and reasoning capabilities. SPIN is based on the fast performance and rich expressivity of SPARQL and allows deterministic reasoning using closed world semantics and unique name assumption while the other technologies in the semantic stack are primarily meant for declarative semantics or the semantics that change relatively slowly. The specifications with dynamic semantics are numerous and can be found in applications such as inferencing of new content, and responding to external stimulus including user-initiated interface actions, time, users' personal profiles, data on a server and condition changes.

We met our objective to provide domain specific alternatives to represent UML-based models of complex conceptual work products. We converted the UML based design of class diagram and state machine to a semantic web model that automatically checks the consistency of data and provides the logical and formal representation of system specifications. We demonstrated that our approach permits to represent dynamic semantics such as the behavioral state machine and can be used for the purpose of model checking of complex and interactive systems. This development enables important applications in model checking of critical and complex systems, a successful verification technique for formal verification of integrative systems.

The workflow modeling for case management was done using an implementation of the Business Process Modeling Notation (BPMN)[30]. We have not yet taken the step to use the conceptual work product to verify the case management system in a model checker, but work to develop a translator for BPMN into the SPIN language for process model checking is nearly complete [31]. Currently there are few tools to formally verify BPMN process models. There is some limited work that looks at choreographies [32] , and other work that turns BPEL, very similar to BPMN, models into Petri Nets but those translations do not include data structures which are critical for conceptual work product verification [33][34][35]. Recent work explores the possibility of translating a subset of BPMN 2.0 into the input language for the SPIN model checker (distinct from the SPIN rule language in this paper) [36]. The SPIN model checker is well suited to verifying that a BPMN model implements a given conceptual work products because of its support for C-like data structures, its native message passing support, and the fact that it is widely deployed with a proven track record of scaling to large models [37]. The SPIN model checker input language lends itself rather naturally to Petri Net like constructs so it is possible to leverage the Petri Net like semantics of BPMN similar to the work converting BPEL to Petri Nets, but with the ability to also include data-objects, data-stores, and message passing to naturally support the BPMN 2.0 standard. Early experiments with the translation are encouraging showing the translation to cover most aspects of the standard somewhat intuitively.

Future work to formally verify if a BPMN process implements a conceptual work product specification is to convert the conceptual work product state diagram into a set of linear temporal logic properties suitable for the SPIN model checker [38]. The properties combined with the BPMN translation to the model checker input language will enable the verification of the entire system. Another approach worth exploring is to translate the conceptual work product state diagram into an automaton representing the language of valid sequences of conceptual work product state transitions and then use that language to search for sequences in the BPMN model that are outside the language (e.g., witness traces to property violations) [39]. Future work will look at the trade-offs between these two ways to use the conceptual work product state diagram in formally verifying if a BPMN accomplishes the intended work in this abstract product.

Finally, this research was funded as part of a larger method to integrate usable health IT with effective and efficient clinical workflows [1]. The method, however, is general to critical, technical interactive systems in many other domains. Verified models of conceptual work products have several valuable enabling roles. They are fundamental requirements that simplify the design of interactive software by providing a precise understanding of the work products it must accomplish as output. They enable model checking to verify the effectiveness of complex systems that will be carried out by complex interactions between people and computers. One of the most difficult design problems for those systems is how to decide the allocation of functionality between people and computers [2]. Model checking makes it meaningful to compare and analyze different verified design options for functionality because they each accomplish equivalent work products. Comparing design for qualities such as cost-effectiveness or usability, is similarly only meaningful if each option has been verified to accomplish equivalent work products. These are key steps that are needed to realize the great potential for interactive computing that will be reliably beneficial to clinical health care and many other critical domains.

## ACKNOWLEDGEMENTS

This research was partially supported by the Agency for Healthcare Research and Quality under Award Number R01HS021233 and by the National Library of Medicine of the National Institutes of Health under Award Number R01LM011829. The content is solely the responsibility of the authors and does not necessarily represent the official views of the funding agency.